\documentclass[12pt]{article}
\usepackage{graphicx}

\font\mybb=msbm10 at 12pt

\font\myeu=eufm10 at 12pt
\def\bb#1{\hbox{\mybb#1}}

\def\frak#1{\hbox{\myeu#1}}
\def\ZZ {\bb{Z}}

\def\g {\frak{g}}

\def\ep{\epsilon}

\newcommand{\beq}{\begin{equation}}
\newcommand{\eeq}{\end{equation}}
\newcommand{\beqa}{\begin{eqnarray}}
\newcommand{\eeqa}{\end{eqnarray}}
\newcommand{\bcen}{\begin{center}}
\newcommand{\ecen}{\end{center}}
\newcommand{\btab}[1]{\begin{tabular}{#1}}
\newcommand{\etab}{\end{tabular}}

\begin{document}
\begin{titlepage}

\setcounter{page}{0}
\begin{flushright}
KEK Preprint 2001-119 \\
hep-th/0109193
\end{flushright}

\vskip 5mm
\begin{center}
{\Large\bf On Asymmetric Orbifolds \\
and the $D=5$ No-modulus Supergravity\\}
\vskip 10mm

{\large
Shun'ya Mizoguchi\footnote{\tt mizoguch@post.kek.jp}
} \\
\vskip 5mm
{\it Institute of Particle and Nuclear Studies \\
High Energy Accelerator Research Organization (KEK) \\
Oho 1-1, Tsukuba, Ibaraki 305-0801, Japan} \\
\end{center}
\vskip 10mm
\centerline{{\bf{Abstract}}}
\vskip 3mm
We examine whether any type II asymmetric orbifolds have 
the same massless spectrum as the dimensional reduction of 
$D=5$ simple supergravity, which, besides the eleven-dimensional
supergravity, is the only known supergravity above four dimensions
with no moduli.
We attempt to construct such models by further twisting the
orbifolds which yield $D=4$, $N=4$ pure supergravity to
find that, unfortunately, none of the models have that spectrum. 
We provide supergravity arguments explaining why this is so.
As a by-product, we list all possible momentum-winding lattices
that give $D=4$, $N=4$ pure supergravity.

\end{titlepage}
\baselineskip=18pt
\setcounter{footnote}{0}
\setcounter{equation}{0}

\section{Introduction}
Besides the eleven-dimensional supergravity, $D=5$ simple 
supergravity is the only known supergravity above four dimensions 
with no moduli, and in fact they have many similarities 
\cite{Cremmer}-\cite{DHJN}. Because $D=5$ simple supergravity contains
no scalar fields and particularly no dilaton, it cannot be realized
as  low-energy theories of any perturbative string compactifications. 
It is rather mysterious why such a supergravity exists 
in only five dimensions and how it arises in string theories.

In view of the similarity of the Lagrangians, one would naturally
speculate that some $D=4$, $N=2$ string compactification might 
lead to the $D=5$ theory in its strong coupling, just as type IIA 
string theory becomes M theory \cite{Townsend,Witten} in this limit.
This idea of finding isolated points in string/M theory moduli space
is an old one and some works based on it can be found in the
literature 
\cite{HMV}-\cite{DH}. The underlying motivation is to find mechanisms
of stabilizing the moduli (see e.g. \cite{Dine}) through the search 
of any `$D=5$ analogue of M theory'.

We will, therefore, look for $D=4$, $N=2$ string compactifications
with  the same massless field content as the dimensional reduction of 
the $D=5$ supergravity. Note that this dimensionally reduced 
theory has only one $N=2$ gravity multiplet and one vector 
multiplet but no hypermultiplet, while any Calabi-Yau or symmetric
orbifold compactifications necessarily contain the universal
hypermultiplet, of which the dilaton is a member, hence it cannot 
be realized by them.
Thus, in this paper we will consider asymmetric
orbifolds 
\footnote{
We concentrate on type II compactifications because 
$D=5$ simple supergravity is obtained \cite{PT} from a consistent
truncation of the $D=11$ supergravity  (and hence the $D=4$ theory 
from IIA theory). } 
\cite{asymmetric}.

Our strategy is the following: We first construct asymmetric
orbifolds with $D=4$, $N=4$ pure supergravity as their low-energy
limits, following ref.\cite{DH}. We then examine if any residual 
symmetries of the invariant lattices fix the shift vector and, 
at the same time, reduce the number of massless vectors to 1/3.
Finally, if there is one, we check if any new massless moduli
appear in the twisted sector. We will show that unfortunately 
{\em none} of the models lead to the desired supergravity in this
framework. We will also give supergravity arguments explaining why
this is so.

Although our attempt is unsuccessful, it provides a no-go 
statement and will be a step toward the understanding the
string-theory origin of the D=5 no-modulus supergravity. As a
by-product, we list all (in this framework) possible momentum-winding
lattices that give $D=4$, $N=4$ pure supergravity by using the
classification of conjugacy classes of Weyl group elements.

In Section 2, we recall the asymmetric orbifold constructions 
of $N=4$ pure supergravity, and give a list of possible
momentum-winding lattices. Section 3 is devoted to the details 
of the constructions. In Section 4, we provide supergravity
arguments suggesting that any orbifold does not seem
likely to realize the desired supergravity as its low-energy
theory. Finally we summarize our conclusions in Section 5.

\section{$N=4$ pure supergravity from asymmetric orbifolds}
We start with a compactification of type II theories on a
six-dimensional torus whose metric is given by the matrix of
some simply-laced semi-simple Lie algebra $\g$. Our convention is
$\alpha'=2$ and the radii $R=1$ so that it is a self-dual torus.
Turning on an appropriate $B$ field, the  Narain lattice 
$\Gamma^{6,6}(\g)={(p_L,p_R)}$
takes the form \cite{Narain}
\begin{eqnarray}
p_L,~p_R\in\Lambda_W(\g),~~~
p_L-p_R\in\Lambda_R(\g),
\label{Narain}
\end{eqnarray}
where $\Lambda_R(\g)$ ($\Lambda_W(\g)$) is the root (weight) lattice 
of a simply-laced (in order for the Cartan matrix to be interpreted 
as a metric) Lie algebra $\g$. The necessary and sufficient condition
for this (with the metric $G_{ij}$ of the torus being the Cartan
matrix) is  that
$(B_{ij}-G_{ij})/2$ have integer entries, where
$i,j$ are the coordinates of the torus. Then $\Gamma^{6,6}(\g)$ is
invariant under the independent actions  on the left- and the
right-lattices of the Weyl reflection group $\cal{W}(\g)$
(T-duality). One can use these automorphisms of the lattice to twist
the model on this special background. We define  the action of $g_L,
g_R\in\cal{W}(\g)$ on a state with definite momenta $|~p_L,p_R>$ as 
\begin{eqnarray}
|~p_L,p_R>\rightarrow e^{2\pi i(p_Lv_L-p_Rv_R)}|~g_Lp_L,g_Rp_R>,
\end{eqnarray}
where, $(v_L,v_R)$ are called the shift vectors. The oscillators 
are transformed similarly.

Let us now construct asymmetric orbifolds whose low-energy limits 
are $N=4$ pure supergravity \cite{Das,CS} following ref.\cite{DH}.
We consider abelian asymmetric orbifolds twisted by a group 
generated by a pair of Weyl group elements $(g_L, g_R)$. To obtain 
an $N=4$ theory whose supersymmetries come only from the 
right-moving sector, we choose $g_R=1$ and $g_L\in SO(6)$ but 
$\notin SU(3)$. Conjugacy classes of Weyl group elements have 
been classified in the mathematical literature 
\cite{Carter,Bouwknegt,LSW}. The statement is that each simple Lie 
algebra $\g$ has a particular set of Weyl group elements known 
as the `primitive elements' (see e.g. \cite{Bouwknegt} for the 
precise definition), and any conjugacy class of Weyl group 
elements of $\g$ corresponds to a primitive element of some 
regular subalgebra of $\g$, that is, the subalgebra whose Dynkin 
diagram is obtained by removing one node from the extended 
Dynkin diagram of $\g$ (or the one by repeating this procedure 
on the Dynkin diagrams so obtained).

In the $\mbox{rank}=6$ case, eigenvalues of a Weyl group 
element $w$ of order $N$ are of the form
\begin{eqnarray}
\{\ep^{r_1},\ep^{r_2},\ep^{r_3},
\ep^{N-r_1},\ep^{N-r_2},\ep^{N-r_3} \}, 
\end{eqnarray}
where $r_i$ and $N-r_i$ are positive integers and 
$\ep=e^{2\pi i/N}$. The condition for $w$ not to lie in $SU(3)$ 
is that any of $r_1\pm r_2\pm r_3\neq 0$ mod $N$. Furthermore,
if $N$ is even, modular invariance require that
the sum $\sum_i r_i$ must be even \cite{Vafa}. An exhaustive search
using  the list of the primitive elements shows that 
these requirements leave only four possible Weyl group elements
listed in Table.  

\begin{center}
\begin{tabular}{|c|c|c|c|}
\hline
Weyl group elements&Eigenvalues&Order&Momentum-winding lattice 
($\g$) 
\\ \hline 
$E_6(a_1)$&$(\ep^1,\ep^2,\ep^4,\ep^5,\ep^7,\ep^8)$&$9$&$E_6$\\
$A_4\oplus A_2$&$(\ep^3,\ep^6,\ep^9,\ep^{12},\ep^5,\ep^{10})$&$15$
&$A_4\oplus A_2$\\
$D_4(a_1)\oplus A_2$&$(\ep^3,\ep^3,\ep^9,\ep^9,\ep^4,\ep^8)$&$12$
&$D_4\oplus A_2$\\
$A_2\oplus A_1\oplus
A_1\oplus A_1\oplus A_1$&$(\ep^2,\ep^4,\ep^3,\ep^3,\ep^3)$&$6$
&$A_2\oplus A_1\oplus
A_1\oplus A_1\oplus A_1$\\
\hline
\end{tabular}
\end{center}
\centerline{Table: List of possible Weyl group elements.
$\ep=e^{2\pi i/N}$, where $N$ is the order}
\centerline{~~~~~~~~~~of the element. 
$a_1$ is a label distinguishing different primitive elements.}
In either case, no larger semi-simple simply-laced Lie algebra 
has the algebra in the last column as its regular subalgebra. 
Therefore, the momentum lattice is uniquely determined for each 
Weyl group element in this table.

Since these twists yield the fields of $D=4$, $N=4$ pure supergravity
already in the untwisted sector, we will take the shift vectors which
are not  in the dual of the invariant lattices \cite{DH} so as to
avoid  extra massless fields from the twisted sector.  Let us examine
each case in some detail. 
\subsection{$E_6(a_1)$}
The first $E_6$ case is a known example \cite{Dixon,DH}. 
In the $A_2\oplus A_2\oplus A_2$ basis  
the metric and the $B$ field are
\begin{eqnarray}
G_{ij}=\left[\begin{array}{cccccc}
2&1&&&&\\
1&2&&&&\\
&&2&1&&\\
&&1&2&&\\
&&&&2&1\\
&&&&1&2
\end{array}
\right],~~~
B_{ij}=\left[\begin{array}{cccccc}
&3&&&&\\
-3&&&&&\\
&&&3&&\\
&&\!\!\!\!\!-3&&&\\
&&&&&3\\
&&&&\!\!\!\!\!-3&
\end{array}
\right].
\end{eqnarray}
We can take 
\begin{eqnarray}\mbox{
\begin{tabular}{ll}
$g_L=(\ep^1,\ep^2,\ep^4,\ep^5,\ep^7,\ep^8)$,&$v_L=0$,
\\
$g_R=1$,&
$v_R=
(\frac19,\frac19,-\frac29;~\frac19,\frac19,-\frac29;~
-\frac19,-\frac19,\frac29)$
\end{tabular}}
\end{eqnarray}
with $\ep=e^{2\pi i/9}$.
The shift vectors are expressed in terms of the weights 
of its regular subalgebra $A_2\oplus A_2\oplus A_2$, 
where the two simple roots of $A_2=SU(3)$ are denoted by 
$(1,-1,0)$ and $(0,1,-1)$ in this notation.
This twist gives a $D=4$, $N=4$ multiplet as the 
massless field already in the untwisted sector. For the 
twisted sector to have no massless fields $v_R$ has been 
so chosen that any of $jv_R$ ($j=1,\ldots,8$) does not 
lie in $\Lambda_W(E_6)$. Since $\frac12 v_R^2=\frac19$, 
the level-matching  condition is satisfied.  

\subsection{$A_4\oplus A_2$}
The next example is the case $A_4\oplus A_2$. 
We found the following two shift vectors:
\begin{eqnarray}
\mbox{
\begin{tabular}{ll}
$g_L=(\ep^3,\ep^6,\ep^9,\ep^{12},\ep^5,\ep^{10})$,&$v_L=0$,
\\
$g_R=1$,&$v_R=(\frac5{15},-\frac4{15},-\frac1{15},0,0~;~
\frac15,-\frac15,0)$
\end{tabular}}
\end{eqnarray}
and 
\begin{eqnarray}
\mbox{
\begin{tabular}{ll}
$g_L=(\ep^3,\ep^6,\ep^9,\ep^{12},\ep^5,\ep^{10})$,&$v_L=0$,
\\
$g_R=1$,&$v_R=(\frac1{15},-\frac2{15},-\frac3{15},\frac4{15},0~;~
0,0,0)$
\end{tabular}
}\end{eqnarray}
with $\ep=e^{2\pi i/15}$.
The shift vectors are again written as vectors in the weight 
spaces. The notation for the $A_2$ piece is the same as above, 
and $(1,-1,0,0,0)$, $(0,1,-1,0,0)$, $(0,0,1,-1,0)$ and 
$(0,0,0,1,-1)$ correspond to the simple roots for the $A_4$ piece. 
Clearly in both cases $15v_R\in\Lambda_R(A_4\oplus A_2)
\subset\Lambda_W(A_4\oplus A_2)$, and  $jv_R$ for any 
$j=1,\ldots,14$ does not belong to the weight lattice.
The level-matching condition is also satisfied. 

\subsection{$D_4(a_1)\oplus A_2$}
We can take
\begin{eqnarray}
\mbox{
\begin{tabular}{ll}
$g_L=(\ep^3,\ep^3,\ep^9,\ep^9,\ep^4,\ep^8)$,&$v_L=0$,
\\
$g_R=1$,&$v_R=(\frac4{12},\frac1{12},\frac1{12},0
~;~\frac1{12},\frac1{12},-\frac2{12})$
\end{tabular}
}\label{D4+A2}
\end{eqnarray}
with $\ep=e^{2\pi i/12}$. The first four entries of $v_R$ is 
a vector in the $D_4$ weight space, where we take 
$(1,-1,0,0)$, $(0,1,-1,0)$, $(0,0,1,-1)$ and $(0,0,1,1)$ as 
its simple roots.

\subsection{$A_2 \oplus A_1\oplus A_1\oplus A_1\oplus A_1$}
Similarly,
\begin{eqnarray}
\mbox{
\begin{tabular}{ll}
$g_L=(\ep^2,\ep^4,\ep^3,\ep^3,\ep^3,\ep^3)$,&$v_L=0$,
\\
$g_R=1$,&$v_R=(\frac16,\frac16,-\frac26~;~
\frac1{6\sqrt2};\frac1{6\sqrt2};\frac1{6\sqrt2};\frac3{6\sqrt2})$
\end{tabular}
}
\label{A2+A1+A1+A1+A1}
\end{eqnarray}
with $\ep=e^{2\pi i/6}$ are a solution. 
The fundamental weight of
$A_1$ is  denoted by $\frac1{\sqrt2}$.
\section{$N=2$ models
by a further twist}

We will now attempt to construct $N=2$ models which have 
a graviton, two vectors and two scalars as their only massless 
bosonic fields. As already mentioned in Introduction, we try 
to construct such models by accompanying a further twist in 
the right sector to reduce the number of vectors (and at the same 
time the number of supersymmetries), while keeping the shift 
vector to avoid extra moduli from the twisted sector. We only 
consider the case where the order of the new right twist is a 
divisor of the order of the left twist since otherwise the order 
of the whole group would change and the level-matching property
of  the original shift vector would be lost.

\subsection{$E_6(a_1)$}
The extended Dynkin diagram of $E_6$ has a $\ZZ_3$ outer 
automorphism, and the shift vector is invariant under the 
$\ZZ_3$ transformation generated by 
\begin{eqnarray}
\left[
\begin{array}{rrr}
0&~1&0\\
0&0&-1\\
-1&0&0
\end{array}
\right],\label{Z3}
\end{eqnarray}
where each block represents the action on one of the
three $A_2$ weight spaces. This belongs to $SU(2)$, and 
hence breaks one half of supersymmetries. Moreover, 
six vectors coming from the NS-NS massless states 
$\psi_{-1/2}^\mu\widetilde{\psi}_{-1/2}^{i}|~\mbox{vac}>$ 
($\mu=2,3$  and $i=4,\ldots,9$ label the noncompact 
and compact coordinates.) are reduced to two by the 
permutation. Thus the twist

\begin{eqnarray}
\mbox{\begin{tabular}{ll}
$g_L=(\ep^1,\ep^2,\ep^4,\ep^5,\ep^7,\ep^8)$,&$v_L=0$,
\\
$g_R=\mbox{eq.}(\ref{Z3})$,&
$v_R=
(\frac19,\frac19,-\frac29;~\frac19,\frac19,-\frac29;~
-\frac19,-\frac19,\frac29)$
\end{tabular}}
\label{newZ9action}
\end{eqnarray}

\noindent
precisely yields the desired massless fields in the 
untwisted sector. However, although the shift vector is
preserved, it turns out that the twisted  sector also have 
extra massless fields. This can been seen as follows:  
Because of the new $\ZZ_3$ right twist, the invariant lattice 
of (\ref{newZ9action}) changed from the original $\Lambda_R(E_6)$ 
to 
\begin{eqnarray}
\{
(v,v,-v)~|~v\in\Lambda_R(A_2)
\}.
\end{eqnarray}
Then its dual lattice becomes
\begin{eqnarray}
\{
\frac13(w,w,-w)~|~w\in\Lambda_W(A_2)
\},
\end{eqnarray}
to which the shift vector $v_R$ belongs. Thus extra 
massless fields arise from the twisted sector.

\subsection{$A_4\oplus A_2$}
The situation is worse. The normal lattice of the shift vector 
has neither $\ZZ_3$ nor $\ZZ_5$ symmetry (Recall that the order 
of the left twist is 15 in this case.), but has only
$\ZZ_2$  symmetry, for both cases found in the last section. Thus the 
untwisted sector has more than two vector fields and the 
model is not the one we wanted to have.

\subsection{The other cases}
A similar analysis shows that neither of the choices (\ref{D4+A2}),
(\ref{A2+A1+A1+A1+A1})
leads to a model with one gravity and one vector multiplet. 
The resulting models have extra massless fields in 
the twisted sector and/or the untwisted sector. This is 
because, again, either the invariant lattices of the $N=4$ models
do not have enough residual symmetries or the new less dense
invariant lattices give rise to extra massless states in the twisted
sector. Although we have explicitly confirmed this only for the shift
vectors given in Section 2, the situation appears to be the same for
the other cases since, after all, choosing a shift vector not lying
in the weight lattice $\Lambda_W(\g)$ is somewhat contradictory to the
demands of preserving enough symmetries.

\section{Supergravity arguments}
Finally, let us consider from the point of view of supergravity
actions why we did not get any asymmetric orbifold model which
realizes the dimensional reduction of $D=5$ simple supergravity. In
general, any type II orbifold compactification has a graviton, a
dilaton and an  anti-symmetric two-form field $B$ as massless fields
coming from  the NS-NS sector. On the other hand, the dimensionally
reduced  bosonic action takes the form \cite{CN} 
\begin{eqnarray}
2\kappa_{4}S_{4}
&=&
\int d^{4}x E^{(4)}\left(
R^{(4)}-\frac32\partial_\mu\rho\partial^\mu\rho
-\frac32\rho^{-2}\partial_\mu A\partial^\mu A
-\frac1{4}\rho^3
B^{(\mbox{\scriptsize KK})}_{\mu\nu}
B^{(\mbox{\scriptsize KK})\mu\nu}
\right.\nonumber\\
&&~~~~~~~~~~~~~~
-\frac3{4}\rho
F^{(4)}_{\mu\nu}
F^{(4)\mu\nu}
\left.-\frac3{4}E^{(4)-1}\epsilon^{\mu\nu\tau\sigma}
F_{\mu\nu}F_{\tau\sigma}A
\right),
\label{D=4}
\end{eqnarray}
where the five-dimensional fields have been 
parameterized as 
\begin{eqnarray}
E^{(5)\hat{\alpha}}_{~\hat{\mu}}&=&
\left[
\begin{array}{cc}
\rho^{-1/2}E^{(4)\alpha}_{~\mu}
&\rho B^{\mbox{\scriptsize (KK)}}_\mu\\
0&\rho
\end{array}
\right],\nonumber\\
A^{(5)}_{\hat{\mu}}&=&\left\{
\mbox{$\begin{array}{l}
A_{\mu} ~~~(\mu=0,\ldots,3),
\\
A_{4}=A,
\end{array}$}\right.
\label{EandA}
\end{eqnarray}
and 
$F^{(4)}_{\mu\nu}=F'_{\mu\nu}+B^{(\mbox{\scriptsize
KK})}_{\mu\nu}A$, $F'_{\mu\nu}=2\partial_{[\mu}A'_{\nu]}$, 
$A'_\mu=A_\mu-B^{\mbox{\scriptsize (KK)}}_\mu A$. $\rho$ is
naturally identified as the dilaton, and therefore the other scalar 
field $A$ must be the dual to the $B$ field if (\ref{D=4}) is a
low-energy  action of an orbifold. However, if the dual
transformation is applied on $A$ (This can be done if we take the
vector $A_\mu$ as fundamental rather than the Kaluza-Klein invariant
one $A'_\mu$.), the dualized action becomes non-polynomial in $A^\mu
A_\mu$ and, unlike type IIA supergravity, is hard to be regarded as a
string effective action.  This is in contrast with $D=4$, $N=4$ pure
supergravity,  which can be successfully dualized \cite{NT} to give a
part of the  heterotic string action. Also, the minimally
coupled $D=4$, $N=2$ supergravity \cite{dWvP} of the type in
ref.\cite{Luciani}, obtained \cite{Das} 
by a consistent truncation of $N=4$ pure supergravity, 
can be easily dualized, in which the two vector fields enter in the
action symmetrically. In our case, however, the two vectors are
asymmetric, carrying different $SO(2)$ charges \cite{MO}. Therefore
the scalar field $A$ cannot be interpreted as an axion.

Where does this scalar come from? We can gain insight into this
question by examining how $D=5$ simple supergravity is obtained 
by a consistent truncation from $D=11$. It is known 
that the eleven-dimensional supergravity (bosonic) action \cite{CJS} 
\begin{eqnarray}
2\kappa_{11}S_{11}
&=&
\int d^{11}x E^{(11)}\left(
R^{(11)}-\frac1{48}F^{(11)}_{MNPQ}F^{(11)MNPQ}
\right.\nonumber\\
&&~~~~~~~~~~~~~~~~
\left.-\frac1{12^4}E^{(11)-1}\epsilon^{M_1\cdots M_{11}}
A^{(11)}_{M_1M_2M_3}F^{(11)}_{M_4\cdots M_7}F^{(11)}_{M_8\cdots
M_{11}}
\right)
\end{eqnarray}
is consistently truncated to $D=5$ simple supergravity 
\begin{eqnarray}
2\kappa_{5}S_{5}
&=&
\int d^{5}x E^{(5)}\left(
R^{(5)}-\frac3{4}
F^{(5)}_{\hat{\mu}\hat{\nu}}F^{(5)\hat{\mu}\hat{\nu}}
\right.\nonumber\\
&&~~~~~~~~~~~~~~
\left.-\frac1{4}E^{(5)-1}\epsilon^{\hat{\mu}_1\cdots \hat{\mu}_{5}}
A^{(5)}_{\hat{\mu}_1}F^{(5)}_{\hat{\mu}_2\hat{\mu}_3}
F^{(5)}_{\hat{\mu}_4\hat{\mu}_5}
\right)
\end{eqnarray}
by setting \cite{PT}
\begin{eqnarray}
E^{(11)A}_{~~M}&=&
\left[
\begin{array}{cc}
E^{(5)\hat{\alpha}}_{~\hat{\mu}}&0\\
0&\delta_{\hat{m}}^{~\hat{a}}
\end{array}
\right],\nonumber\\
A^{(11)}_{MNP}&=&
\left\{\mbox{$\begin{array}{l}
A_{\hat{\mu}56}=
A_{\hat{\mu}78}=A_{\hat{\mu}9~\!10}=A_{\hat{\mu}},
\\
0~~~\mbox{otherwise,}
\end{array}$}\right.
\end{eqnarray}
where $M,N,\ldots$ and $A$ are the eleven-dimensional curved 
and local-Lorentz indices, and $\hat{\mu},\hat{\nu}\ldots$ 
and $\hat{\alpha}$ are the corresponding five-dimensional 
indices. $\hat{m}$ and $\hat{a}$ are those of the flat
six-dimensional internal space $T^6$. 
The reduced $D=4$ action is obtained by further parameterizing
the five-dimensional fields as (\ref{EandA}).

Now if the fifth direction $M=4$ is thought of as 
tangent to the $S^1$ direction which is compactified to  
yield type IIA theory, then $\rho$ is certainly the dilaton, 
while $A=A_{456}=A_{478}=A_{49~\!10}$ are the components of the 
NS-NS two-form $B$ field with both indices tangent to $T^6$.
The $B_{\mu\nu}$ components are truncated and do not appear 
in the spectrum. This is the reason why the dualization of $A$ 
does not give a neat result. Another difference from the massless 
spectrum of the orbifold is that both of two vector fields are
in the R-R sector, while what we have trying to restore are 
the vectors in the NS-NS sector. Of course, in general one can
think of U dualities which interchange NS-NS and R-R, but an 
explicit investigation shows that no $E_7$ element maps 
the dimensionally reduced supergravity to the minimally coupled 
one. Thus we conclude that there is no reason to expect that the 
reduction of $D=5$ supergravity to
$D=4$ is  realized by the orbifolds that we have considered. 
Also, even though there exists a totally different asymmetric
orbifold which is not related to N=4 pure supergravity but 
has one $N=2$ gravity and one vector multiplet as its massless 
fields, it would not be related to the five-dimensional theory 
in any limit.

\section{Conclusions}
We have tried to construct, by further twisting  
the $N=4$ pure supergravity models, asymmetric
orbifolds whose massless fields are the same as the
dimensional reduction of $D=5$ simple supergravity. 
We have found no example of models, and argued that this is in some
sense natural because the second scalar is not the axion and the two
vectors should come from the R-R sector.

We have seen that it is hard to fix all moduli but one $N=2$ 
vector multiplet; it cannot be achieved by Calabi-Yau
compacitifications, nor by orbifolds. The string-theory origin of
$D=5$ simple supergravity still remains obscure. However, the
arguments in the last section indicate another possibility of 
finding $D=5$ simple supergravity in string theories: We have seen
that the
$B_{\mu\nu}$ components with  four-dimensional spacetime indices are
truncated, and we know models in which this truncation occurs: The
orientifolds. It would be interesting to investigate whether any
orientifold model realizes the
$D=4$ spectrum, and in case there is such a model, whether it has a
decompactifying limit to $D=5$.

\end{document}